\documentclass[floats,floatfix,showpacs,amssymb,prd,twocolumn,superscriptaddress,nofootinbib]{revtex4-1}
\usepackage{adjustbox}
\usepackage{anyfontsize}
\usepackage{amssymb}
\usepackage{amsmath}
\usepackage{amsfonts}
\usepackage{amsbsy}
\usepackage{txfonts}

\usepackage{subfigure}
\usepackage{gensymb}
\usepackage{soul}
\usepackage{wrapfig}

\def\namedlabel#1#2{\begingroup
    #2%
    \def\@currentlabel{#2}%
    \phantomsection\label{#1}\endgroup
}

\usepackage{mathrsfs}

\usepackage{bm}

\newcommand{\dG}{\Delta G/G_\text{N}}

\usepackage[utf8]{inputenc}

\usepackage{graphicx}
\usepackage{epsfig}
\usepackage{epstopdf}

\usepackage[usenames,dvipsnames]{xcolor}

\usepackage{verbatim,mathtools,needspace,enumitem,etoolbox}

\definecolor{linkcolor}{rgb}{0.0,0.3,0.5}

\usepackage[unicode, colorlinks=true, linkcolor=linkcolor, citecolor=linkcolor, filecolor=linkcolor,urlcolor=linkcolor, pdfusetitle]{hyperref}

\usepackage{epstopdf}

\usepackage{bm}
\usepackage{dcolumn}

\usepackage{longtable}
 \usepackage[normalem]{ulem}

\definecolor{romared}{RGB}{142,0,28}
\hypersetup{colorlinks=true,
            citecolor=romared,
            linkcolor=romared,
            urlcolor=romared}
\begin{document}

\title{Screened fifth forces lower the TRGB-calibrated Hubble constant too}

\author{Harry Desmond}
\email{harry.desmond@physics.ox.ac.uk}
\affiliation{Astrophysics, University of Oxford, Denys Wilkinson Building, Keble Road, Oxford OX1 3RH, UK}

\author{Jeremy Sakstein}
\email{sakstein@hawaii.edu}
\affiliation{Department of Physics \& Astronomy, University of Hawai'i, 2505 Correa Road, Honolulu, Hawai'i, 96822, USA}


\raggedbottom

\begin{abstract}
The local distance ladder measurement of the Hubble constant requires a connection between geometric distances at low redshift and Type Ia supernovae in the Hubble flow, which may be achieved through either the Cepheid period--luminosity relation or the luminosity of the Tip of the Red Giant Branch (TRGB) feature of the Hertzsprung--Russell diagram. Any potential solution to the Hubble tension that works by altering the distance ladder must produce consistency of both the Cepheid and TRGB $H_0$ calibrations with the CMB. In this paper we extend our models of screened fifth forces \cite{D19} to cover the TRGB framework. A fifth force lowers TRGB luminosity, so a reduction in inferred $H_0$ requires that the stars that calibrate the luminosity---currently in the LMC---are on average less screened than those that calibrate the supernova magnitude. We show that even under pessimistic assumptions for the extinction to the LMC, full consistency with \textit{Planck} can be achieved for a fifth force strength in unscreened RGB stars $\sim$0.2 that of Newtonian gravity. This is allowed by the comparison of Cepheid and TRGB distance measurements to nearby galaxies. Our results indicate that the framework of \cite{D19} is more versatile than initially demonstrated, capable of ameliorating the Hubble tension on a second front.  
\end{abstract}

\date{\today}

\maketitle

\section{Introduction}
\label{sec:intro}

\begin{figure*}[ht]
    \centering
    \includegraphics[width=0.5\textwidth]{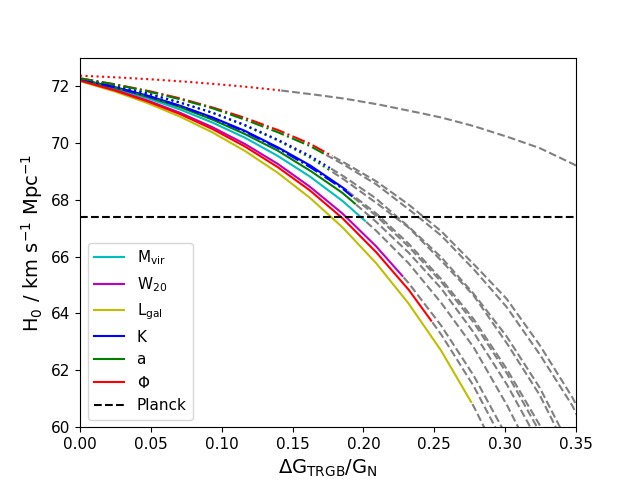}\hfill
    \includegraphics[width=0.5\textwidth]{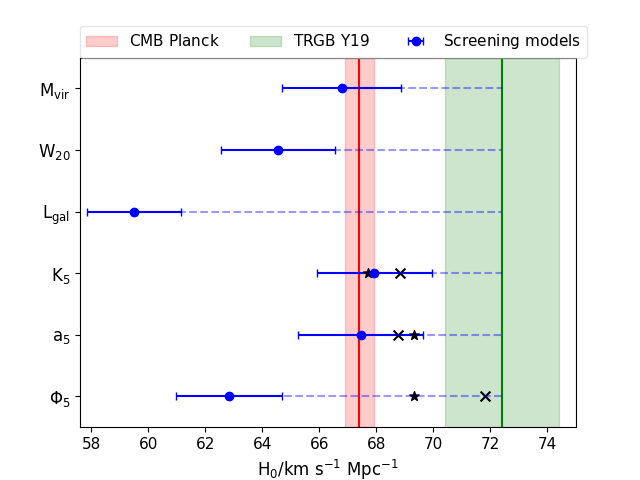}
    \caption{\emph{Left:} True TRGB $H_0$ values in our models as a function of $\Delta G_\text{TRGB}/G_\text{N}$. The lines correspond to different screening proxies: halo virial mass $M_\text{vir}$, HI linewidth $W_{20}$, galaxy luminosity $L_\text{gal}$, and environmental curvature $K$, acceleration $a$ and Newtonian potential $\Phi$. The lines become dashed gray where they enter $5\sigma$ tension with the Cepheid vs TRGB distance test (Sec.~\ref{sec:tests}). For the environmental proxies the 0.5, 5 and 50 Mpc apertures are shown by dotted, solid and dotdashed lines respectively. \emph{Right:} True $H_0$ at the maximum values of $\Delta G_\text{TRGB}/G_\text{N}$ allowed by the distance test, separately for each proxy (points with errorbars). By using a smaller $\Delta G_\text{TRGB}/G_\text{N}$ any value of $H_0$ can be achieved along the horizontal dashed blue lines. The red and green vertical lines and shaded regions show the best-fit $H_0$ value, and its $1\sigma$ uncertainty, from \textit{Planck} and Y19 respectively. For the environmental proxies $\Phi$, $a$ and $K$ we show results for a 5 Mpc aperture, but indicate also the result of using a 0.5 or 50 Mpc aperture with black crosses and stars respectively (the errorbars are similar). Most screening models are fully able to reconcile Y19 with \textit{Planck} while maintaining self-consistency of the distance ladder.}
    \label{fig:results}
\end{figure*}

The $\Lambda$CDM model has been hugely successful at explaining cosmological observations, but, as data has become more precise, tensions have begun to appear. The most statistically significant currently is the Hubble tension: low-redshift probes of the universe's present expansion rate $H_0$ prefer higher values than the cosmic microwave background (CMB). A debate is now underway as to whether this discrepancy signals new physics at play in the universe (e.g. see \cite{Verde, HHG} and references therein).

In previous work \cite[hereafter D19]{D19}, we showed that extensions of General Relativity (GR) that incorporate a screened fifth force include a mechanism for lowering the Hubble constant measured through the local distance ladder. The idea is that differences in the masses or environmental densities of the galaxies used to calibrate the Cepheid period--luminosity relation (PLR) vs those that use the PLR to determine supernova Ia (SN) luminosity can generate different fifth force strengths within the corresponding sets of Cepheids. We described how a fifth force shifts the PLR from the Newtonian relation, and hence biases $H_0$ high if the calibration Cepheids are screened while some fraction of the cosmological ones are not. In some models this reduces the tension with \textit{Planck} from $4.4\sigma$ to $\sim$2$\sigma$. This improvement is limited by the consistency of distance measurements to galaxies though either the Cepheid or TRGB method, as we discuss further below.

Cepheids currently provide the most precise measurement of $H_0$ in the local universe \cite{R16,R18,R19} (hereafter R16, R18, R19), but a growing number of other low-redshift probes have produced $H_0$ values different to the \textit{Planck} result of $67.4 \pm 0.5$ km s$^{-1}$ Mpc$^{-1}$. These include calibration of the distance ladder by means of the TRGB feature \cite[hereafter F19 and F20]{F19,F20} or Mira variables \cite{Miras_2} rather than Cepheids, distance measurements from megamasers \cite{megamasers} or strong lensing time delays \cite{Suyu,Wong}, $\gamma$-ray attenuation \cite{Dominguez}, cosmic chronometers \cite{CC1,CC2} and baryon acoustic oscillations \cite{BAO1,BAO2,BAO3}. The fact that most of these prefer values above 67.4 km s$^{-1}$ Mpc$^{-1}$ reduces the likelihood that the Hubble tension is due to systematic error in low-redshift methods, and instead suggests that new physics may invalidate direct comparison of values measured at either end of the universe's expansion history. Any new physical mechanism would ideally be capable of reconciling the values of $H_0$ achieved by all methods; here we undertake this task for the fifth force model with regard to the difference between the TRGB measurement and \textit{Planck}.

The presence of an unscreened fifth force in the hydrogen-burning shells of RGB stars reduces the TRGB luminosity, causing the distance to the corresponding galaxy to be overestimated if GR is assumed (D19). As a universal shift in the TRGB absolute magnitude has no net effect on inferred distances, a modification to the inferred $H_0$ requires different screening on average in the calibration and cosmological samples. The current TRGB calibration of the distance ladder uses the Large Magellanic Cloud (LMC) to anchor the TRGB absolute $I$-band magnitude $M_I$ because its distance is well known by geometrical methods \cite{DEB}. Our models in this paper work by unscreening the LMC---so that $M_I$ is biased high---but screening a significant fraction of the SN host galaxies that the TRGB feature is used to estimate the distance to. This causes their distances to be larger than inferred, and hence $H_0$ to be overpredicted.

While the only fundamental requirement for reducing the TRGB $H_0$ is that the LMC be unscreened, satisfying local tests of GR requires that the Milky Way (MW; at least at the position of the Solar System) be screened. Since the MW is used to calibrate the Cepheid PLR this is required to reduce the Cepheid $H_0$ too, and in D19 we set the threshold such that the MW is just screened.\footnote{The Cepheid PLR is also calibrated with the maser distance to NGC 4258, but in almost all screening models this is more screened than the MW. We discuss the LMC calibration of the PLR further in Sec.~\ref{sec:discussion}.} We use the same screening thresholds here, so that our Cepheid results are unaffected. It is important to note however that, as discussed in detail in D19, some specific screening mechanisms such as chameleon are constrained by other probes to such an extent that they are not able to alter $H_0$ appreciably.

The $\Lambda$CDM measurement of $H_0$ through the TRGB feature is currently the subject of debate in the literature. F19 report $H_0 = 69.8 \pm 0.8$ (stat) $\pm 1.7$ (sys) km s$^{-1}$ Mpc$^{-1}$, or $69.6 \pm 0.8$ (stat) $\pm 1.7$ (sys) km s$^{-1}$ Mpc$^{-1}$ using a slightly updated model for the extinction to the LMC (F20). Conversely, \cite[hereafter Y19]{Y19} argue that F19 significantly overestimate the LMC extinction, leading to $M_I$ larger by about 0.08 mag and hence a higher $H_0$: $72.4 \pm 2$ km s$^{-1}$ Mpc$^{-1}$. The discrepancies with \textit{Planck} \cite{planck} of the F20 and Y19 results are $1.1\sigma$ and $2.4\sigma$ respectively. Thus while neither calibration is strongly discrepant with \textit{Planck}, they add to the Hubble tension when combined with the independent Cepheid measurements. There may also be an environment-dependence of TRGB luminosity in GR due to mass loss and the consumption of nearby planets \cite{Jimenez:2020hsd}. We will remain agnostic as to the true TRGB $H_0$ in GR. However, since the F20 result is already essentially consistent with the CMB we will prioritise the Y19 value, simply to show that even in that case our models can restore consistency. This requires a fifth force strength $\dG \simeq 0.2$ for RGB stars.

The plan of this paper is as follows. In Sec.~\ref{sec:method} we describe our model for the TRGB feature in the presence of a fifth force, and our methodology for propagating this into the $H_0$ constraint. In Sec.~\ref{sec:results} we derive the screening properties of the F19 sample under a variety of models, expand on the Cepheid vs TRGB distance test as a means of bounding the fifth force strength in RGB stars, and use this information to determine the maximum possible effect of our screening models on $H_0$. The main results are shown in Fig.~\ref{fig:results} (see Sec.~\ref{sec:H0_effect}). Sec.~\ref{sec:discussion} describes some caveats of our analysis, discusses our results in the broader context of the $H_0$ tension and suggests avenues for further work. Sec.~\ref{sec:conc} concludes.

\section{Methodology}
\label{sec:method}

Our methodology is similar to D19, to which we refer the reader for further detail. The overall aim is to calculate a modified constraint on $H_0$ as calibrated by the TRGB method, starting from an existing GR calibration (e.g. F20 or Y19) but taking into account the action of a screened fifth force. We use the F19 sample comprising 18 SNe in 15 galaxies, plus the LMC as calibrator (see F19 table 3).

\subsection{Effect of a fifth force on TRGB}

A fifth force with range much larger than the size of a star obeys the inverse-square law and therefore behaves like a boost to Newton's constant \cite{Burrage:2017qrf}; we describe its strength by $\dG \equiv (G-G_\text{N})/G_\text{N}$. In D19 we ran simulations of TRGB stars under modified gravity using a modified version of the \textit{Modules for Experimentation in Stellar Astrophysics} (\textsc{mesa}) code \cite{Paxton:2010ji,Paxton:2013pj,Paxton:2015jva,Paxton:2017eie} and found a fitting function for their distance as a function of $G$:
\begin{equation}\label{eq:d_trgb}
\frac{\bar{D}}{D} = 1.021 \left(1-0.04663 \left(1+\frac{\Delta G}{G_\text{N}}\right)^{8.389} \right)^{1/2} \equiv K(\Delta G).
\end{equation}
The overbar denotes the ``true'' value, whereas $D$ is the distance that would be inferred in GR (i.e. $G=G_N$). For general $\Delta G>0$, $\bar{D} < D$. We will assume that all unscreened objects experience the same $\Delta G$, as determined in Sec.~\ref{sec:tests}.

\subsection{Determination of screening proxies}

The phenomenon of screening implies that only some objects feel the fifth force, while others experience regular gravity. Screening generally occurs in dense environments, but the exact quantity that governs it (the ``screening proxy'') varies between models. We investigate the same set of screening proxies as we did in D19: i) galaxy luminosity $L_\text{gal}$ (measured by visual magnitude $M_V$), ii) galaxy internal velocity (measured by HI linewidth $W_{20}$), iii) halo virial mass $M_\text{vir}$, and iv) environmental gravitational field. The latter is quantified by the Newtonian potential $\Phi$, acceleration $a$ or curvature $K$ sourced by mass within 0.5, 5 or 50 Mpc of the test point. The only proxy of D19 that we do not include here is the local dark matter density, $\rho_\text{DM}$. This is because the locations of the stars sourcing the TRGB feature within their dark matter halos are not known.

It is worth bearing in mind that a fifth force can alter both $M_V$ and $W_{20}$ in unscreened galaxies. While $W_{20}$ has the relatively mild scaling $W_{20} \propto (G/G_\text{N})^{1/2}$, $L_\text{gal} \propto (G/G_\text{N})^3$ \cite{Davis} and may therefore be a factor $\sim$2 larger than in screened but otherwise identical galaxies for $G/G_\text{N} \sim 1.2$ as considered here. The galaxies in our sample span two orders of magnitude in $L_\text{gal}$, so differences in $L_\text{gal}$ are driven by differences in stellar mass regardless of the screening of main sequence stars. We have checked explicitly using data from the \textit{NASA-Sloan Atlas}\footnote{\url{http://www.nsatlas.org/data}} that if $L_\text{gal}$ is considered a proxy for the more physical screening indicator $M_*$, the screening misclassification fraction is increased only slightly when the $G$-dependence of $L_\text{gal}$ in unscreened galaxies is included. The majority of theories screen main sequence stars anyway however, so this effect is unlikely to be present at all.

For each galaxy in the F19 sample, $M_V$ and its uncertainty is taken from the \textit{NASA Extragalactic Database},\footnote{\url{https://ned.ipac.caltech.edu/}} and $W_{20}$ from the \textit{Extragalactic Distance Database}.\footnote{\url{http://edd.ifa.hawaii.edu/}}$^-$\footnote{$M_V$ is not listed for M66, or $W_{20}$ for N4526, N1316, N1404 or the LMC. We fill these data in using linear regression between $M_V$ and $W_{20}$ for the galaxies where both are measured, and assign the conservative uncertainties $\Delta M_V = 0.4$ for M66, $\Delta W_{20} = 20$ km/s for the LMC, and $\Delta W_{20} = 30$ km/s for N4526, N1316 and N1404.} $M_\text{vir}$ is determined by using the best-fitting halo abundance matching model of \cite{Lehmann} to link $M_V$ to halos in the \textsc{DarkSky}-400 $N$-body simulation \cite{DarkSky}. Uncertainties are given as the standard deviations of the $M_\text{vir}$ distribution for each galaxy derived by repeating the abundance matching 200 times. $\Phi$, $a$ and $K$, and their uncertainties, are derived from the screening maps of \cite{Desmond} evaluated at the positions of the galaxies.

\subsection{Effect on $H_0$}

Eq.~\ref{eq:d_trgb} applies for a fixed value of TRGB absolute magnitude $M_I$. However, if the LMC calibrates the unscreened value, $M_I^\text{unscr}$, the screened value is given by
\begin{equation}
M_I^\text{scr} = M_I^\text{unscr} + 5\log(K(\Delta G_\text{LMC})).
\end{equation}
An unscreened galaxy (for $\Delta G = \Delta G_\text{LMC}$) has $M_I = M_I^\text{unscr}$ whereas a screened galaxy has $M_I = M_I^\text{scr}$. Since $K<1$, $M_I^\text{scr} < M_I^\text{unscr}$ so that distances to screened galaxies are underestimated by a factor $K(\Delta G)$ when their TRGB magnitudes are assumed to equal $M_I^\text{unscr}$, i.e. $\bar{D} = D/K(\Delta G)$.

$H_0$ is derived from the Hubble diagram of SNe with absolute magnitude $M_\text{SN}$ calibrated at low redshift using either Cepheids or TRGBs. In both cases $H_0$ is calculated as
\begin{equation}\label{eq:H0_1}
H_0 = 10^{M_\text{SN}/5 + 5 + a_B} \; \text{km s}^{-1} \text{Mpc}^{-1},
\end{equation}
where $a_B$ is the normalization of the magnitude--redshift relation of the cosmological SNe sample, for which we use the R16 value $a_B = 0.713 \pm 0.002$. Writing this in terms of the SN apparent magnitude $m_\text{SN}$ and distance $D$, the $H_0$ value implied by galaxy $i$ is
\begin{equation}\label{eq:H0_2}
H_{0,i} = 10^{m_\text{SN,i}/5 + 6 + a_B} \: \frac{\text{pc}}{D_i} \; \text{km s}^{-1} \text{Mpc}^{-1}.
\end{equation}
Using $\bar{D_i} = D_i/K(\Delta G_i)$, the true $H_0$ value is given by
\begin{equation}\label{eq:H0_3}
\bar{H}_{0,i} = H_{0,i} K_i \: ;
\end{equation}
i.e. $\bar{H}_{0,i} < H_{0,i}$. As in D19, we assume that in the standard analysis each galaxy implies the same final value of $H_0$, i.e. $H_{0,i} = H_0^\text{GR}$ where $H_0^\text{GR}$ is the overall $H_0$ value at $\dG=0$. This could be the F20 or Y19 best-fit value (or similar); we use the Y19 result $H_0^\text{GR} = 72.4 \pm 2.0$ km s$^{-1}$ Mpc$^{-1}$ because it is more discrepant with \textit{Planck}.

For each galaxy, we estimate the uncertainty in the $H_0$ estimate as
\begin{equation}\label{eq:delta_H0}
\Delta H_{0,i}^2 = \ln(10)^2 \: (\Delta m_\text{SN,i}^2/25 + \Delta a_B^2) + (\Delta D_i/D_i)^2,
\end{equation}
where $\Delta m_\text{SN,i}$ and $\Delta D_i$ are the measurement uncertainties on $m_\text{SN,i}$ and $D_i$ respectively. These are taken from F19 table 3.\footnote{For galaxies with multiple SNe we take an average magnitude uncertainty.} We assume the same fractional uncertainty on our modified $H_0$ constraint, i.e. $\Delta \bar{H}_{0,i}/\bar{H}_{0,i} = \Delta H_{0,i}/H_{0,i}$. We average the $\bar{H}_{0,i}$ values over the F19 galaxies weighted by their uncertainties to derive an overall best-fit estimate of $\bar{H}_0$:
\begin{equation}\label{eq:H0_4}
\bar{H}_0 = H_0^\text{GR} \frac{\sum_i K_i {\Delta \bar{H}_{0,i}^{-2}}}{\sum_i {\Delta \bar{H}_{0,i}^{-2}}}.
\end{equation}
We then scatter this by $\Delta H_0^\text{GR} \: \bar{H}_0/H_0^\text{GR}$ to account for the uncertainty on $H_0^\text{GR}$.

The final source of uncertainty in $\bar{H}_0$ derives from the degree of screening of the galaxies, due to uncertainties in their proxy values. To take this into account we repeat the above procedure 10,000 times, in each case randomly scattering the proxy values by their uncertainties. This alters which galaxies in the cosmological sample are screened, and hence $K_i$ and $\bar{H}_{0,i}$. Our final estimate for $\bar{H}_0$, $\hat{H}_0$, is the mean of these 10,000 $\bar{H}_0$ values, and the uncertainty $\Delta \hat{H}_0$ is their standard deviation. We quantify the resulting tension with \textit{Planck} as $\sigma_{H_0} \equiv (\hat{H}_0 - H_\text{0,CMB})/(\Delta \hat{H}_0^2+\Delta H_\text{0,CMB}^2)^{1/2}$.

\section{Results}
\label{sec:results}

In this section we give the screening properties of the F19 galaxies, upgrade the Cepheid vs TRGB distance test of D19 for limiting the possible fifth force strength, and determine the TRGB $H_0$ values our models are able to produce.

\subsection{Screening properties of TRGB hosts}
\label{sec:screening_props}

In Fig.~\ref{fig:proxies} (Appendix~\ref{sec:app1}) we show the screening proxies over the F19 galaxies as determined by the procedure of Sec.~\ref{sec:method}. The errorbars show the minimal width enclosing $68\%$ of the model realisations, the red line with shaded errorbar shows the value for the LMC, and the dashed magenta line shows the MW value. We also show in dashed green the average value over the R16 Cepheid hosts (cf. D19 fig. 3). Galaxies above the magenta line for a given model are screened in our formalism and therefore contribute to lowering the TRGB $H_0$. We see that the TRGB hosts are on average more screened than the Cepheid hosts, following the expectation that Cepheids tend to be found in young late-type galaxies and TRGBs in older ellipticals (e.g. F19). This is helpful for allowing our screening scenario to reduce both the Cepheid and TRGB $H_0$. The proxy values for the F19 galaxies are also listed in Tables~\ref{tab:galprops1} and~\ref{tab:galprops2}.

It is important to assess the region of parameter space in which our framework effectively reduces $H_0$ under both Cepheid and TRGB calibrations of the distance ladder. Although we perform the calculations for the case in which the MW is only just screened, our mechanism works for any screening threshold between the MW and LMC values. This region may be seen on a proxy-by-proxy basis in Fig.~\ref{fig:proxies}: for some screening models such as the one based on $L_\text{gal}$ this allows almost an order of magnitude range in the proxy, while for others such as those based on the large-scale gravitational field the viable window is small. This provides insight into the kind of full screening theory that would be most effective at lowering the local $H_0$ through our mechanism.

The second column of Table~\ref{tab:H0} shows the unscreened fraction of SN host galaxies for each proxy. There is a large variation between screening models, but in most cases the majority of hosts are screened. The effect of a screening model on $H_0$ is increased by a smaller unscreened fraction for the SN hosts, and by a larger allowed $\Delta G/G_N$ as described below.

\subsection{Maximum $\Delta G/G_{\rm N}$: the Cepheid--TRGB distance test}
\label{sec:tests}

D19 identified the observable $(D_\text{Ceph}-D_\text{TRGB})/D_\text{TRGB} \equiv \Delta D/D$ as useful for providing an upper bound on fifth force strength. This is because a GR-based analysis underestimates $D_\text{Ceph}$ and overestimates $D_\text{TRGB}$ in the presence of a fifth force, so the consistency of $D_\text{Ceph}$ and $D_\text{TRGB}$ on a galaxy-by-galaxy basis places a limit on $\Delta G/G_N$. We use the same 51 galaxies from NED-D as D19, but make three modifications to the test. First, we standardise the TRGB distances to a common value of the absolute magnitude $M_I$, either $-4.047$ (F20) or $-3.963$ (Y19), for consistency with those used in the distance ladder. The distances are calibrated to $M_I = -4.1$ by default, and this correction has a negligible impact on the results. Second, we allow a separate $\dG$ for the Cepheids and TRGB in a given galaxy. This is motivated by the environment-dependence of screening mechanisms and the fact that Cepheids and RGB stars occupy different regions within their host halos. There is likely some further spread within each population, but separate $\Delta G_\text{TRGB}/G_\text{N}$ and $\Delta G_\text{Ceph}/G_\text{N}$ is a good first approximation given the uncertainties involved. Phenomenologically, the result is that $\Delta G_\text{TRGB}/\Delta G_\text{N}$ may exceed $\Delta G_\text{Ceph}/G_\text{N}$ when Cepheid cores are unscreened because the TRGB is less sensitive to the fifth force than Cepheids. The expectation for $\Delta D/D$ is then
\begin{align}\label{eq:dd}
\left\langle \frac{\Delta D}{D} \right\rangle = &\left(1+\frac{\Delta G_\text{Ceph}}{G_\text{N}}\right)^{-\frac{A+2B}{4}} \\ \nonumber
& \times 1.021 \left(1-0.04663 \left(1+\frac{\Delta G_\text{TRGB}}{G_\text{N}}\right)^{8.389} \right)^{1/2} - 1,
\end{align}
\noindent where $A$ and $B$ describe the contributions to the distance modification from unscreening the Cepheid envelope and core respectively (cf. D19 eq. 6).  
Finally, we now determine the values of the screening proxies explicitly over the 51 galaxies in the dataset, rather than using the average unscreened fractions from the R16 (or F19) samples as we did before. This enables us to pin down more precisely the maximum $\dG$ values that the NED-D dataset allows. The proxies are determined by the same means as in Sec.~\ref{sec:method}. Galaxies with missing data in NED or EDD are excluded on a proxy-by-proxy basis; this removes 3 galaxies for $M_V$ and $M_\text{vir}$ and 8 for $W_{20}$.

We describe the expected distribution of $\Delta D/D$ values using a Gaussian likelihood:
\begin{equation}
\ln \mathcal{L}\left(\frac{\Delta D}{D}_\text{obs} \: \bigg\rvert \: \frac{\Delta G}{G_\text{N}}, \sigma_\text{n}\right) = -\frac{(\frac{\Delta D}{D}_\text{obs} - \langle \frac{\Delta D}{D} \rangle)^2}{2 \sigma_\text{tot}^2} - \frac{1}{2} \ln(2 \pi \sigma_\text{tot}^2),
\end{equation}
where $\sigma_\text{tot}^2 \equiv \sigma_{\Delta D/D}^2 + \sigma_\text{n}^2$, $\sigma_{\Delta D/D}$ is propagated from the quoted uncertainties on $D_\text{Ceph}$ and $D_\text{TRGB}$, and $\sigma_\text{n}$ is an additional noise term accounting for possible astrophysical contributions to $\Delta D$ not captured by the measurement uncertainty. In screened galaxies we of course expect $\langle \Delta D/D\rangle = 0$.

As discussed in D19, in some screening models Cepheid cores are screened (e.g. chameleons), while in others, such as the baryon--dark matter interaction model, they are not. We find that although the constraint on $\Delta G_\text{Ceph}/G_\text{N}$ depends sensitively on whether or not Cepheid cores are unscreened (it is much tighter if they are), the constraint on $\Delta G_\text{TRGB}/\Delta G_\text{N}$ does not. This is illustrated in Fig.~\ref{fig:corner}, which shows the 3D corner plot of the constraints on $\Delta G_\text{Ceph}/G_\text{N}$, $\Delta G_\text{TRGB}/\Delta G_\text{N}$ and $\sigma_n$ for an example case in which 30\% of the galaxies are unscreened, for Cepheid cores screened (left) or unscreened (right). The third column of Table~\ref{tab:H0} lists the maximum values of $\Delta G_\text{TRGB}/\Delta G_\text{N}$ for each of the proxies from this test, showing that values up to $\sim$0.2 are allowed. As in D19, we use $5\sigma$ limits due to the possibility of systematic errors in the test stemming from combining inhomogeneous data sets and treating them as independent. These values may be statistically allowed in a fully self-consistent comparison of Cepheid and TRGB distances, which we leave for future work.

\subsection{Effect on $H_0$}
\label{sec:H0_effect}

The remaining columns of Table~\ref{tab:H0} give the best-fit values of $H_0$, their uncertainties and the resulting discrepancies with \textit{Planck}. These are illustrated in Fig.~\ref{fig:results}: the left panel shows $\hat{H}_0$ as a function of $\Delta G_\text{TRGB}/\Delta G_\text{N}$ for each screening model (the lines become dashed gray where they exceed the corresponding $\Delta G_\text{TRGB}/\Delta G_\text{N}$ limits), while the right panel shows $\hat{H}_0$ and its $1\sigma$ uncertainty at the maximum allowed $\Delta G_\text{TRGB}/\Delta G_\text{N}$. We start from the Y19 rather than F20 $H_0$ value simply because this sets a harder task for our models. We see that the majority of our models are nevertheless able to achieve full consistency with \textit{Planck}---indeed, some require significantly smaller $\Delta G_\text{TRGB}/\Delta G_\text{N}$ than they are permitted by the Cepheid--TRGB distance test. The luminosity-based model is the most effective because in that model all bar one of the SN calibrators in the F19 sample are screened (Fig.~\ref{fig:proxies} top left).


\section{Discussion}
\label{sec:discussion}

In this section we provide more detail on systematic uncertainty in our analysis and discuss ways in which the fifth force scenario may be further investigated in the future.

An important role in our framework is played by the statistical consistency of distance measurements using Cepheids and the TRGB to galaxies where both are observed. This is the most stringent test we have identified for bounding modifications to the strength of gravity within these stars. Cepheid and TRGB distances have previously been compared (\cite{Jain:2012tn} and F19) but the comparison has not been used to make detailed statements about possible systematics or new physics in their relation. F19 note that the dispersion of $D_\text{Ceph}-D_\text{TRGB}$ is larger than the errorbars, leading them to suggest that the latter may be underestimated. We quantify this with the astrophysical noise parameter $\sigma_n$, which in support of F19 we find to be clearly greater than 0. Our model does however effectively apportion extra error equally to both $D_\text{Ceph}$ and $D_\text{TRGB}$, whereas F19 argue that the extra uncertainty is likely to be mainly in $D_\text{Ceph}$. We also assume that both the measurement errors and extra noise term are Gaussian. It is therefore possible that our use of this test is affected by systematic error, and our constraints should be revisited in light of future more precise measurements and/or better understanding of the uncertainties involved. Were our scenario correct, we would expect ultimately to find TRGB distances normalised higher than Cepheid distance in galaxies with low mass or in low density regions.

While our decision to set the screening thresholds in our models to the MW values ensures consistency with the results on Cepheid-calibrated $H_0$ in D19, the details of how the fifth force must function to lower $H_0$ in the Cepheid vs TRGB cases has the potential to strongly constrain our scenario. In particular, if the TRGB luminosity is calibrated (as currently) with the LMC, the RGB stars of the LMC must be unscreened for our mechanism to work. However, the Cepheids that calibrate the PLR must be screened. These are mainly within the MW and NGC 4258, calibrated by parallax and a water maser respectively, although SH0ES also considers Cepheids in the LMC with distances calibrated from detached eclipsing binaries (R16). In the simplest scenario where degree of screening is common to an entire galaxy, one would expect that if the TRGB feature in the LMC is unscreened then the Cepheids would be too, and hence that the SH0ES pipeline using only the LMC as distance anchor  would infer a significantly lower value of $H_0$, closer to the ``true'' CMB value. This does not appear to be the case (R19). A strengthening of the LMC-only SH0ES result would force us to consider the possibility that RGB stars in the LMC are unscreened while Cepheids are screened. This could possibly derive from their different positions within the LMC halo: Cepheids are typically located nearer the centres of galaxies than RGB stars, in regions of higher density. The precise details of this are model-dependent. A more stringent consistency test could also be devised from a sizeable sample of galaxies with simultaneous TRGB, Cepheid and SN measurements.

The next milestone in the measurement of $H_0$ by the TRGB method will be the calibration of $M_I$ using stars in the MW with parallaxes measured by \textit{Gaia} (F19, F20). If these stars are screened due to the denser environment of the MW relative to the LMC, our model would predict that the resulting $H_0$ value would be significantly lower than is currently attained. In fact, if some of the SN hosts' TRGBs are unscreened but all of those calibrating $M_I$ are screened, we would expect the best-fit $H_0$ to lie \emph{below} $H_0^\text{CMB}$ due to our screening modification acting in the opposite direction. In general, differences between $H_0$ values derived with different calibrators provide important information on our theory.

In D19 and \cite{rho_DM} we identified the baryon--dark matter interaction model, with a screening mechanism governed by the local dark matter density $\rho_\text{DM}$, as particularly promising for lowering $H_0$ without conflicting with other observations. To include this here we would require a model for the distribution of RGB stars within their host halos. This is not currently available, but may become so with more precise future observations. For now we simply note that we expect this model to give results within the range covered by the screening proxies we do consider.

We have now shown that the action of a screened fifth force has the potential to reconcile the local distance ladder calibrated using either Cepheids or TRGBs with the CMB. There remain however other low-redshift measurements of $H_0$ that indicate a Hubble rate higher than \textit{Planck}. The most precise of these is the H0LiCOW measurement using strong lensing time delays of distant quasars, which currently yields $H_0 = 73.3^{+1.7}_{-1.8}$ km s$^{-1}$ Mpc$^{-1}$ \cite{Wong}. We would expect the strong lenses around which the time delays accumulate to be predominantly screened due to their large mass, so that our models would not appreciably alter this result. It may be possible to extend the models to include modifications to lensing in high density regions, but more likely our scenario would have to rely on systematic error within the H0LiCOW pipeline (e.g. \cite{Kochanek, Blum, Kochanek:2020crs}). Should other methods independently achieve strong inconsistency with \textit{Planck}, the fifth force framework would need to be extended to them too.

\section{Conclusion}
\label{sec:conc}

Most proposed resolutions of the Hubble tension based on new physics alter the universe's pre-recombination expansion history. In past work we have suggested a post-recombination solution: a modification to the physics governing the stars that calibrate the local distance ladder. In particular, the fifth forces that are expected to arise in theories beyond GR can shift the normalisation of the Cepheid period--luminosity relation, introducing a bias in the inferred $H_0$ value if the calibration and SN host galaxies are differently screened \cite{D19}. Here we show that the same models can lower the Hubble constant inferred from the TRGB calibration of the distance ladder if the stars used to calibrate the TRGB magnitude (in our case in the LMC) are unscreened, but some of those in SN host galaxies are not. We find that fifth force strengths $\Delta G_\text{TRGB}/\Delta G_\text{N} \simeq 0.2$ are able to `correct' the TRGB $H_0$ value derived by \cite{Y19} to the \textit{Planck} value. These $\Delta G_\text{TRGB}/\Delta G_\text{N}$ values are allowed by the consistency of Cepheid and TRGB distances to common galaxies, which is the most stringent test we have devised for this scenario. If \cite{F20} uses a more accurate extinction model for the LMC than \cite{Y19}, then even smaller values $\Delta G_\text{TRGB}/G_\text{N} \simeq 0.1$ are required.

While more work is required to further constrain the strength of gravity in RGB and Cepheid stars, and extend the framework to other low-redshift probes of $H_0$, our results suggest that screened fifth forces---generic in extended gravity and dark energy models---may offer a solution to the Hubble tension.

\bigskip

{\it Acknowledgements:} 

We thank Bhuvnesh Jain for many enlightening and important discussions.

HD is supported by St John's College, Oxford, and acknowledges financial support from ERC Grant No. 693024 and the Beecroft Trust.


\bibliography{ref_TRGB}

\appendix

\begin{table*}
  \centering
  \small\addtolength{\tabcolsep}{-5pt}
    \begin{tabular}{|l|c|c|c|c|c|c|c|c|}
      \hline
      Name & RA/Dec (J2000 \degree) & $\mu$ & $\Delta \mu$ & $\Delta m_\text{SN}$ & $M_V$ & $\Delta M_V$ & $W_{20}/\text{km/s}$ & $\Delta W_{20}/\text{km/s}$\\ 
      \hline      
    \rule{0pt}{3.5ex}
      LMC & 80.89 / -69.76 & 18.48 & 0.02 & --- & -18.30 & 0.20 & 77 & 20\\
    \rule{0pt}{3.5ex}
      MW & --- & 0 & 0 & --- & -20.60 & 0.50 & 169 & 50\\
    \rule{0pt}{3.5ex}
      M101 & 210.80 / 54.35 & 29.08 & 0.04 & 0.03 & -21.30 & 0.20 & 194 & 5\\
    \rule{0pt}{3.5ex}
      M66 & 170.06 / 12.99 & 30.22 & 0.04 & 0.07 & -22.40 & 0.40 & 381 & 5\\
    \rule{0pt}{3.5ex}
      M96 & 161.69 / 11.82 & 30.31 & 0.04 & 0.06 & -22.50 & 0.20 & 366 & 7\\
    \rule{0pt}{3.5ex}
      N4536 & 188.61 / 2.19 & 30.96 & 0.05 & 0.04 & -22.97 & 0.20 & 353 & 6\\
    \rule{0pt}{3.5ex}
      N4526 & 188.51 / 7.70 & 31.00 & 0.07 & 0.04 & -22.30 & 0.20 & 342 & 30\\
    \rule{0pt}{3.5ex}
      N4424 & 186.8 / 9.42 & 31.00 & 0.06 & 0.06 & -19.20 & 0.20 & 95 & 5\\
    \rule{0pt}{3.5ex}
      N1448 & 56.13 / -44.64 & 31.32 & 0.06 & 0.04 & -21.99 & 0.23 & 414 & 7\\
    \rule{0pt}{3.5ex}
      N1365 & 53.40 / -36.14 & 31.36 & 0.05 & 0.03 & -23.00 & 0.20 & 404 & 5\\
    \rule{0pt}{3.5ex}
      N1316 & 50.67 / -37.21 & 31.46 & 0.04 & 0.027 & -23.69 & 0.17 & 573 & 30\\
    \rule{0pt}{3.5ex}
      N1404 & 54.72 / -35.59 & 31.42 & 0.05 & 0.046 & -22.38 & 0.13 & 352 & 30\\
    \rule{0pt}{3.5ex}
      N4038 & 180.47 / -18.87 & 31.68 & 0.05 & 0.15 & -21.84 & 0.19 & 294 & 8\\
    \rule{0pt}{3.5ex}
      N5584 & 215.60 / -0.39 & 31.82 & 0.10 & 0.05 & -21.00 & 0.20 & 215 & 5\\
    \rule{0pt}{3.5ex}
      N3021 & 147.74 / 33.55 & 32.22 & 0.05 & 0.05 & -21.20 & 0.20 & 291 & 8\\
    \rule{0pt}{3.5ex}
      N3370 & 161.77 / 17.27 & 32.27 & 0.05 & 0.05 & -21.20 & 0.20 & 287 & 12\\
    \rule{0pt}{3.5ex}
      N1309 & 50.53 / -15.40 & 32.50 & 0.07 & 0.04 & -20.8 & 0.20 & 161 & 6\\
    \hline
    \end{tabular}
  \caption{Properties of the F19 TRGB sample (plus the MW). $\mu$ (distance modulus to galaxy) and $\Delta m_\text{SN}$ (uncertainty in SN apparent magnitude) are transcribed from F19, galaxy visual absolute magnitude $M_V$ is obtained from the Nasa Extragalactic Database (NED) and galaxy HI linewidth $W_{20}$ is obtained from the Extragalactic Distance Database (EDD).}
  \label{tab:galprops1}
\end{table*}

\section{Gravitational properties of TRGB hosts and effects on $H_0$}
\label{sec:app1}

\begin{figure*}
  \centering
  \includegraphics[width=0.49\textwidth]{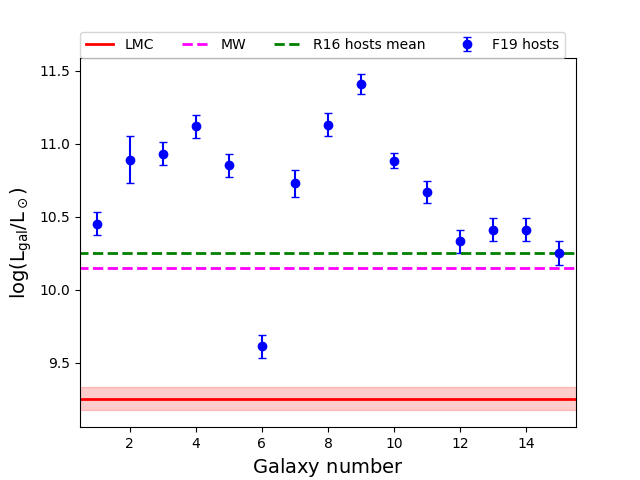}
  \includegraphics[width=0.49\textwidth]{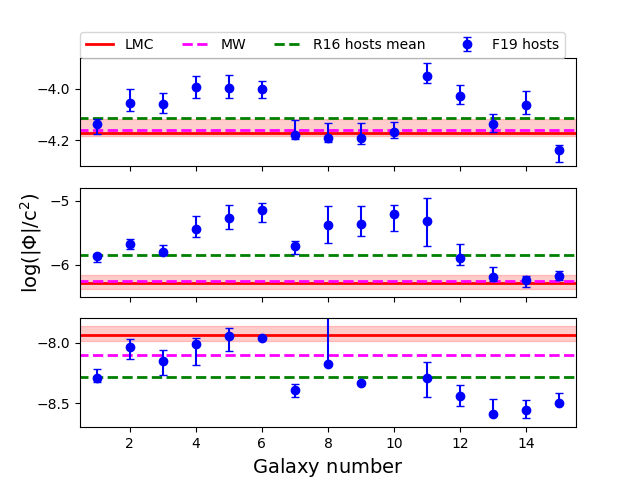}
  \includegraphics[width=0.49\textwidth]{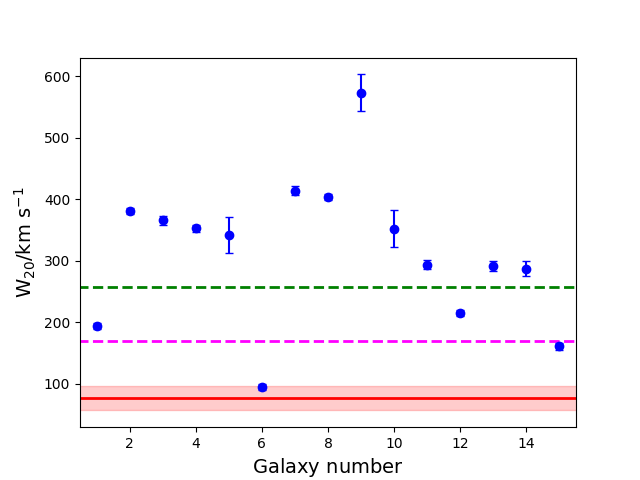}
  \includegraphics[width=0.49\textwidth]{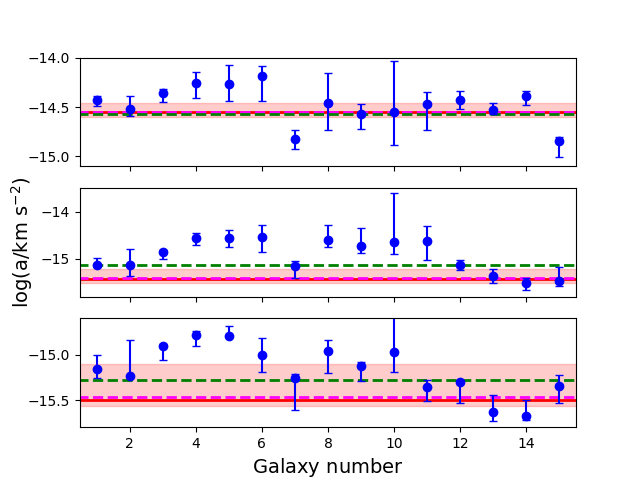}
  \includegraphics[width=0.49\textwidth]{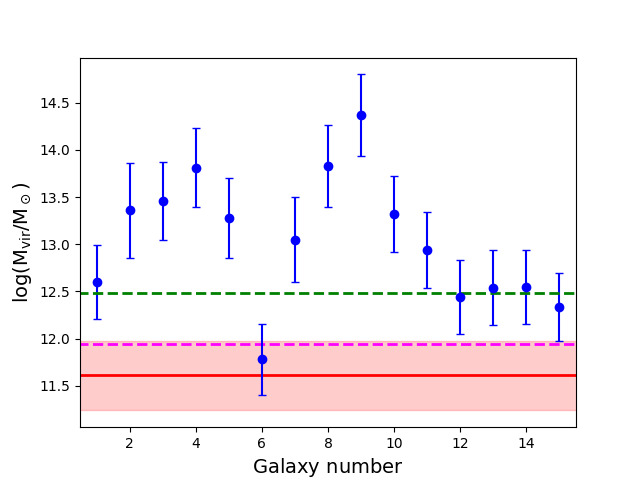}
  \includegraphics[width=0.49\textwidth]{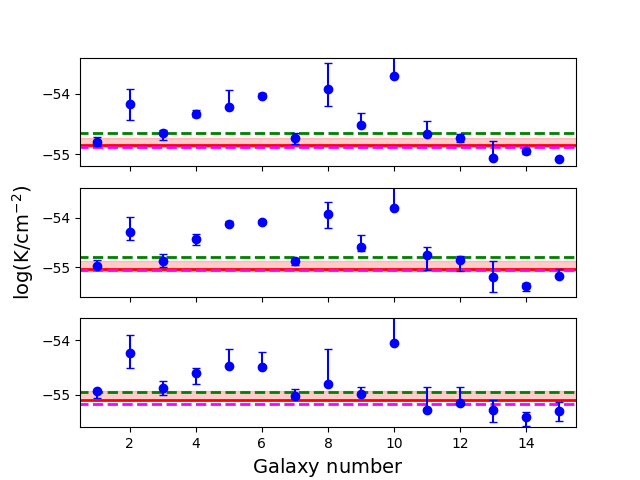}
  \caption{The screening proxies of $V$-band luminosity, velocity dispersion of hydrogen gas, halo virial mass, and external potential, acceleration and curvature over the SN host galaxies in the F19 TRGB sample. The LMC and MW are shown separately by the horizontal red and magenta lines, and the green dashed line shows the average over the R16 Cepheid hosts. Galaxies lying above the magenta line are considered screened (we suppress the MW errorbar for visual clarity, but see D19 fig. 3). The ordering of the galaxies is as in Table~\ref{tab:galprops1} (starting with M101). The subpanels of the environmental screening plots (right column) correspond to the three different distances out to which we include contributions from masses: $0.5$ Mpc (lower), $5$ Mpc (middle) and $50$ Mpc (upper). The asymmetric errorbars indicate the minimal width enclosing $68\%$ of the Monte Carlo model realisations.}
  \label{fig:proxies}
\end{figure*}

\begin{figure*}[ht]
    \centering
    \includegraphics[width=0.475\textwidth]{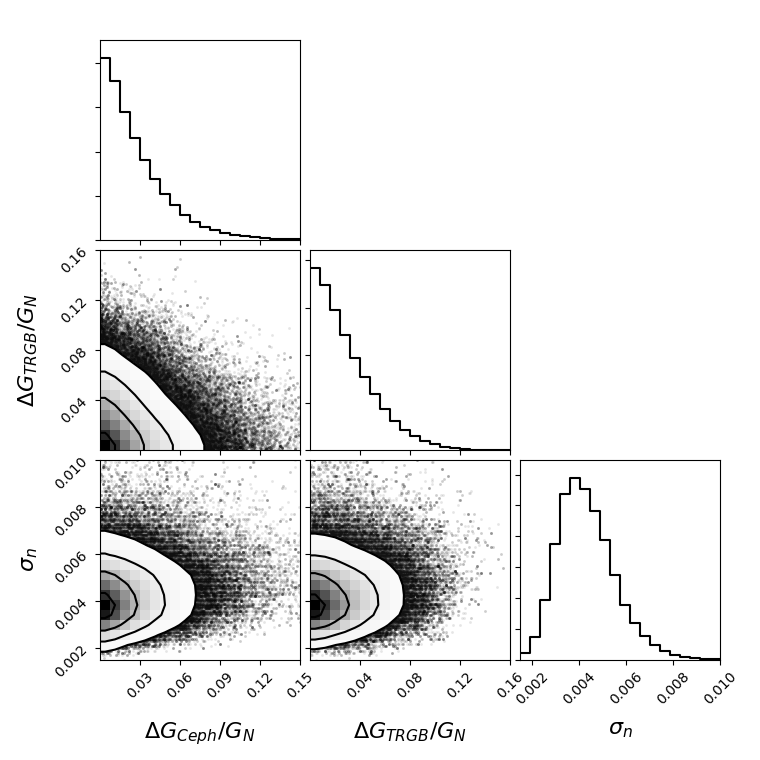}\hfill
    \includegraphics[width=0.475\textwidth]{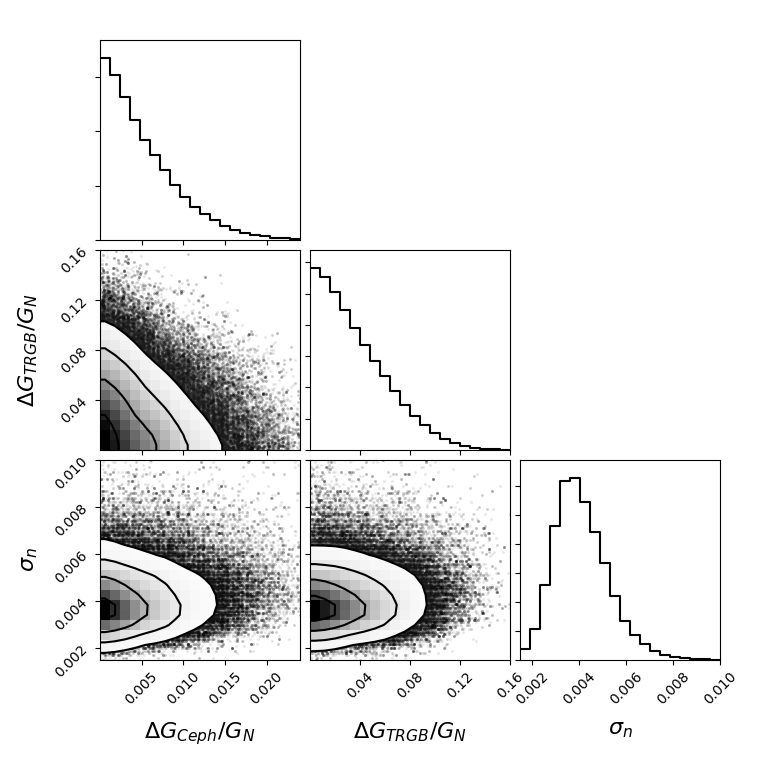}
    \caption{Corner plots for $\Delta G/G_N$ for Cepheids and TRGB, and additional noise $\sigma_n$ in the comparison between Cepheid and TRGB distances. These particular constraints assume $30\%$ of the sample is unscreened. \emph{Left:} Cepheid cores screened, \emph{Right}: Cepheid cores unscreened.}
    \label{fig:corner}
\end{figure*}

Tables~\ref{tab:galprops1} and~\ref{tab:galprops2} list the galaxy and halo properties of the F19 sample respectively, determined by the methods of Sec.~\ref{sec:method}. The proxy values are plotted along with their uncertainties in Fig.~\ref{fig:proxies}. Table~\ref{tab:H0} shows the effects of the models on $\hat{H}_0$, $\Delta \hat{H}_0$ and $\sigma_{H_0}$ for the maximum $\Delta G_\text{TRGB}/\Delta G_\text{N}$ allowed at $5\sigma$ by the Cepheid vs TRGB distance test. Illustrative corner plots for the constraints derived from this test are shown in Fig.~\ref{fig:corner}, for a galaxy unscreened fraction of 0.3 and separately for Cepheid cores screened (left) or unscreened (right).

\begin{table*}[hb]
  \begin{center}
  \small\addtolength{\tabcolsep}{-5pt}
    \begin{tabular}{|l|c|c|c|c|c|c|c|c|}
      \hline
      Name & $\log(R/\text{kpc})$ & $\Delta \log(R/\text{kpc})$ & $\log(c)$ & $\Delta \log(c)$ & $\log(V/\text{km/s})$ & $\Delta \log(V/\text{km/s})$ & $\log(M/M_\odot)$ & $\Delta \log(M/M_\odot)$\\ 
      \hline      
    \rule{0pt}{3.5ex}
      LMC & 2.28 & 0.12 & 1.07 & 0.30 & 1.98 & 0.12 & 11.61 & 0.37\\
    \rule{0pt}{3.5ex}
      MW & 2.37 & 0.12 & 1.07 & 0.27 & 2.24 & 0.12 & 11.94 & 0.37\\
    \rule{0pt}{3.5ex}
      M101 & 2.61 & 0.13 & 0.97 & 0.23 & 2.31 & 0.13 & 12.60 & 0.39\\
    \rule{0pt}{3.5ex}
      M66 & 2.86 & 0.17 & 0.90 & 0.18 & 2.56 & 0.17 & 13.36 & 0.50\\
    \rule{0pt}{3.5ex}
      M96 & 2.90 & 0.14 & 0.87 & 0.18 & 2.60 & 0.14 & 13.46 & 0.41\\
    \rule{0pt}{3.5ex}
      N4536 & 3.02 & 0.14 & 0.84 & 0.16 & 2.72 & 0.14 & 13.81 & 0.42\\
    \rule{0pt}{3.5ex}
      N4526 & 2.84 & 0.14 & 0.90 & 0.19 & 2.54 & 0.14 & 13.28 & 0.42\\
    \rule{0pt}{3.5ex}
      N4424 & 2.34 & 0.13 & 1.07 & 0.27 & 2.04 & 0.13 & 11.78 & 0.38\\
    \rule{0pt}{3.5ex}
      N1448 & 2.76 & 0.15 & 0.95 & 0.20 & 2.46 & 0.15 & 13.05 & 0.45\\
    \rule{0pt}{3.5ex}
      N1365 & 3.02 & 0.14 & 0.84 & 0.16 & 2.72 & 0.14 & 13.83 & 0.43\\
    \rule{0pt}{3.5ex}
      N1316 & 3.20 & 0.14 & 0.80 & 0.16 & 2.90 & 0.14 & 14.37 & 0.43\\
    \rule{0pt}{3.5ex}
      N1404 & 2.85 & 0.13 & 0.90 & 0.18 & 2.55 & 0.13 & 13.32 & 0.40\\
    \rule{0pt}{3.5ex}
      N4038 & 2.72 & 0.13 & 0.96 & 0.20 & 2.43 & 0.13 & 12.94 & 0.40\\
    \rule{0pt}{3.5ex}
      N5584 & 2.56 & 0.13 & 1.02 & 0.23 & 2.26 & 0.13 & 12.44 & 0.39\\
    \rule{0pt}{3.5ex}
      N3021 & 2.59 & 0.13 & 0.99 & 0.23 & 2.29 & 0.13 & 12.54 & 0.40\\
    \rule{0pt}{3.5ex}
      N3370 & 2.60 & 0.13 & 0.99 & 0.23 & 2.30 & 0.13 & 12.55 & 0.39\\
    \rule{0pt}{3.5ex}
      N1309 & 2.52 & 0.12 & 1.01 & 0.22 & 2.22 & 0.12 & 12.34 & 0.36\\
      \hline
    \end{tabular}
  \caption{Halo properties of the galaxy sample, estimated by abundance matching to $M_V$ (see Sec.~\ref{sec:method}). Size $R$, rotation velocity $V$ and enclosed mass $M$ are evaluated at the virial radius.}
  \label{tab:galprops2}
  \end{center}
\end{table*}

\begin{table*}[]
  \begin{center}
  \small\addtolength{\tabcolsep}{-5pt}
    \begin{tabular}{|c|c|c|c|c|c|}
      \hline
      Proxy & Unscreened fraction & $\dG$ & $\hat{H}_0$ / km s$^{-1}$ Mpc$^{-1}$ & $\Delta \hat{H}_0$ / km s$^{-1}$ Mpc$^{-1}$ & $\sigma_{H_0}$\\ 
      \hline
\rule{0pt}{3.5ex}
      --- & --- & 0 & 72.4 & 2.0 & 2.4\\
      \hline      
\rule{0pt}{3.5ex}
      $\Phi_{0.5}$ & 0.85 & 0.14 & 71.8 & 2.0 & 2.1\\
      \hline
\rule{0pt}{3.5ex}
      $\Phi_{5}$  & 0.10 & 0.25 & 62.8 & 1.9 & -2.4\\
      \hline
\rule{0pt}{3.5ex}
      $\Phi_{50}$ & 0.36 & 0.18 & 69.3 & 2.2 & 0.9\\
      \hline
\rule{0pt}{3.5ex}
      $a_{0.5}$ & 0.38 & 0.18 & 68.8 & 2.3 & 0.6\\
      \hline
\rule{0pt}{3.5ex}
      $a_{5}$ & 0.27 & 0.19 & 67.4 & 2.2 & 0.0\\
      \hline
\rule{0pt}{3.5ex}
      $a_{50}$ & 0.43 & 0.17 & 69.3 & 2.1 & 0.9\\
      \hline
\rule{0pt}{3.5ex}
      $K_{0.5}$ & 0.40 & 0.18 & 68.8 & 2.1 & 0.7\\
      \hline
\rule{0pt}{3.5ex}
      $K_{5}$ & 0.30 & 0.19 & 67.9 & 2.0 & 0.3\\
      \hline
\rule{0pt}{3.5ex}
      $K_{50}$ & 0.27 & 0.19 & 67.7 & 2.0 & 0.1\\
      \hline
\rule{0pt}{3.5ex}
      $L_\text{gal}$ & 0.04 & 0.28 & 59.5 & 1.6 & -4.6\\
      \hline
\rule{0pt}{3.5ex}
      $W_{20}$ & 0.14 & 0.23 & 64.5 & 2.0 & -1.4\\
      \hline
\rule{0pt}{3.5ex}
      $M_\text{vir}$ & 0.20 & 0.20 & 66.8 & 2.1 & -0.3\\
      \hline
    \end{tabular}
  \caption{The effect of various screening proxies on the value of $H_0$ inferred from the TRGB-calibrated local distance ladder, assuming the maximum $\Delta G_\text{TRGB}/\Delta G_\text{N}$ at the $5\sigma$ limit of that allowed by the Cepheid vs TRGB distance test (separately for each proxy). The first row indicates the Y19 result. As described in the text, the threshold value for screening under a given proxy is given by the MW's value for that proxy. $\sigma_{H_0}$ is the discrepancy of the resulting $H_0$ constraint with the CMB. Most models are able to achieve full consistency with \textit{Planck}.}
  \label{tab:H0}
  \end{center}
\end{table*}

\end{document}